# STATISTICAL CHALLENGES IN THE ANALYSIS OF COSMIC MICROWAVE BACKGROUND RADIATION

By Paolo Cabella and Domenico Marinucci

*University of Rome Tor Vergata and University of Rome Tor Vergata*

An enormous amount of observations on Cosmic Microwave Background radiation has been collected in the last decade, and much more data are expected in the near future from planned or operating satellite missions. These datasets are a goldmine of information for Cosmology and Theoretical Physics; their efficient exploitation posits several intriguing challenges from the statistical point of view. In this paper we review a number of open problems in CMB data analysis and we present applications to observations from the WMAP mission.

## 1. Introduction.

1.1. *Cosmological background.* Cosmology is now developing into a mature observational science, with a vast array of different experiments that yield datasets of astonishing magnitude and nearly as great challenges for theoretical and applied statisticians. Datasets are now available on a large variety of different phenomena, but the leading part in cosmological research has been played over the last 15 years by the analysis of Cosmic Microwave Background (CMB) radiation, an area which has already led to Nobel Prizes for Physics in 1978 and in 2006.

The nature of CMB can be loosely explained as follows [see, e.g., Dodelson (2003) for a textbook account]. According to the standard cosmological model, the Universe that we currently observe originated approximately 13.7 billion years ago in a very hot and dense state, in what of course is universally known as the Big Bang. Neglecting fundamental physics in the first fractions of seconds, we can naively imagine a fluid state where matter was completely ionized, that is, the kinetic energy of electrons was much stronger than the electrical attraction of protons, so that no stable atomic nuclei could form. It is a consequence of quantum principles that a free electron has a much larger *cross-section* than when it is bound in a nucleus; loosely speaking, as a consequence, the probability of interactions between photons and electrons is so high that the mean free path of the former was very short and the Universe was consequently "opaque." As the Universe expands, the mean energy content decreases, that is, the fluid of matter and radiation cools down; the mean kinetic energy of the electrons decreases as well until it reaches a critical value where it is no longer sufficient to compensate the electromagnetic attraction of the protons; stable (and neutral) hydrogen atoms are then formed. This change of state occurs at the so-called "age of recombination," which is currently reckoned to have taken place $3.7 \times 10^5$ years after the Big Bang, that is, when the Universe had only the 0.003% of its current age. At the age of recombination, the probability of interactions became so small that, as a first approximation, photons could start to travel freely. Neglecting second order effects, we can assume they had no further interaction up to the present epoch.

The remarkable consequence of this mechanism is that the Universe is embedded in a uniform radiation that provides pictures of its state nearly $1.37 \times 10^{10}$ years ago; this is exactly the above-mentioned CMB radiation. The existence of CMB was predicted by G.Gamow in a series of papers in the forties; it was later discovered fortuitously by Penzias and Wilson in 1965—for this discovery they earned the Nobel Prize for Physics in 1978. For several years further experiments were only able to confirm the existence of the radiation, and to test its adherence to the Planckian curve of blackbody emission, as predicted by theorists. A major breakthrough occurred with NASA satellite mission COBE, which was launched in 1989 and publicly released the first full-sky maps of radiation in 1992; for these maps Smoot and Mather earned the Noble Prize for Physics in 2006 [Smoot et al. (1992)].

The nature of these maps deserves further explanation. CMB is distributed in remarkably uniform fashion over the sky, with deviations in the order of $10^{-4}$ with respect to the mean value (corresponding to 2.731 Kelvin degrees). The attempts to understand this uniformity have led to very important developments in cosmology, primarily the inflationary scenario which now dominates the theoretical landscape. Even more important, though, are the tiny fluctuations around this mean value, which provided the seeds for stars and galaxies to form out of gravitational instability. Measuring and understanding the nature of these fluctuations has then been the core of an enormous amount of experimental and theoretical research. In particular, their stochastic properties yield a goldmine of information on a variety of extremely important issues on astrophysics and cosmology, and on many problems at the frontier of fundamental physics.

To mention just a few of these problems, we recall the issues concerning the matter content of the Universe, its global geometry, the existence and nature of (nonbaryonic) *dark matter*, the existence and nature of *dark energy*, which is related to Einstein's cosmological constant, and many others. The next experimental landmark in CMB analysis followed in 2000, when two balloon-borne experiments, BOOMERANG and MAXIMA, yielded the first high-resolution observations on small patches of the sky (less than 10° squared). These observations led to the first constraints on the global geometry of the Universe, which was found to be (very close to) Euclidean. Another major breakthrough followed with the 2003, 2007 and 2008 data releases from another NASA satellite experiment, *WMAP* (the data are publicly available on the web site <http://lambda.gsfc.nasa.gov/>). Such data releases yielded measurement of the correlation structure of the random field up to a resolution of about 0.22 degrees, that is, approximately 30 times better than *COBE* (7–10 degrees). Another major boost in data analysis is expected from the ESA satellite mission *Planck*, which is now scheduled to be launched on October 31, 2008; data releases for the public are expected in the following 3–5 years. Planck is planned to provide datasets of nearly $5 \times 10^{10}$ observations, and this will allow to settle many open questions with CMB temperature data. New challenging questions are expected to arise at a faster and faster pace over the next decades; for instance, Planck will provide high quality for so-called polarization data, which will set the agenda for the experiments to come. Polarization data can be viewed as *tensor-valued*, rather than scalar, observations—that is, what we observe are not measurements of a scalar quantity such as the temperature, but random quadratic forms. As such, this entails an entirely new field of statistical research, which is still in its infancy and will not be discussed in the present paper.

Our aim here is to provide a review of statistical issues arising in CMB data analysis, with many examples of applications of statistical procedures to real data from the *WMAP* experiment. Some of the empirical results we provide are new, as detailed below. The plan of the paper is as follows: in Section 2 we review very briefly some background material on map-making, component separation and spectral representations for the CMB data sets. For brevity's sake, we do not provide many details other than the material which is essential for our following discussion. In Section 3 we are concerned with angular power spectrum estimation, and we discuss procedures to deal with relevant practical questions such as the presence of observational noise and/or missing observations. In Section 4 we present some tools to test for Gaussianity and/or isotropy of CMB radiation: we focus, in particular, on harmonic methods such as the bispectrum, techniques based on differential geometry such as the local curvature, and spherical wavelets (with the so-called Spherical Mexican Hat approach). Concerning the latter, we stress that many other possible approaches to wavelets on the sphere exist, which have been successfully applied to various parts of cosmological and astrophysical research: nevertheless, the field is still extremely active and very much open for research (in particular, the derivation of the stochastic properties of wavelets procedures is still at the very beginning). Finally, we collect in the Appendix some background mathematical material which we considered necessary for a better understanding of our proposals.

## 2. Some preliminary issues.

2.1. *Map-making and component separation.* To understand more precisely the nature of the statistical issues involved, we need to introduce some more formalization. As explained above, CMB can be viewed as the single realization of a random field on the surface of the sphere, that is, for each $x \in S^2$, $T(x)$ is a random variable on a probability space. Observations are provided by means of electromagnetic detectors (so-called radiometers and/or bolometers) which measure fluxes of incoming radiations (i.e., photons) on a range of different frequencies. For instance, the above mentioned *WMAP* experiment is based upon 16 detectors, centered at frequencies 40.7, 60.8 and 93.5 GHz, which are labeled the Q, V and W band, respectively. The forthcoming ESA mission *Planck* will be based upon 70 channels ranging from 30 GHz to 857 GHz. As the satellites scan the sky, observations are collected as a vector time series, the number of observations being in the order of $10^9$ for *WMAP* and $5 \times 10^{10}$ for Planck. A first issue then relates to the construction of spherical maps starting from the Time Ordered Data (TOD) provided by the satellite; this is the so-called *map-making* challenge; see, for instance, Keihanen, Kurki-Suonio and Poutanen (2005) and De Gasperis et al. (2005). For brevity's sake, we shall provide only the basic framework, and refer to the literature for more details. In short, we can assume that in each of the $p$ channels we actually observe

$$O_i(x) = T(x) + F_i(x) + N_i(x), \qquad i = 1, \ldots, p, x \in S^2;$$

here, $T(\cdot)$ denotes the CMB signal, $F_i(x)$ denotes the so-called foreground emissions by galactic and extragalactic sources of noncosmological nature (for instance, galaxies, quasars, intergalactic dusts and others), and $N_i(x)$ instrumental noise. The crucial point to be understood is that the dependence across the different frequency channels of CMB emission is known, and it is different from the pattern followed by other sources: this capital property makes *component separation* possible and allows the construction of filtered maps [see. e.g., Patanchon et al. (2005) and the references therein]. More precisely, a clear prediction from theoretical physics, confirmed to amazing accuracy from the very first experiments [Smoot et al. (1992)], is that CMB radiation should follow the Planckian curve of blackbody radiation, that is, radiation is distributed across frequencies $\nu_i$, $i = 1, \ldots, p$ according to the function

$$R(\nu; x) = \frac{8\pi h v^3}{c^3} \frac{1}{e^{-hv/k_B T(x)} - 1}, \tag{1}$$

Q band map

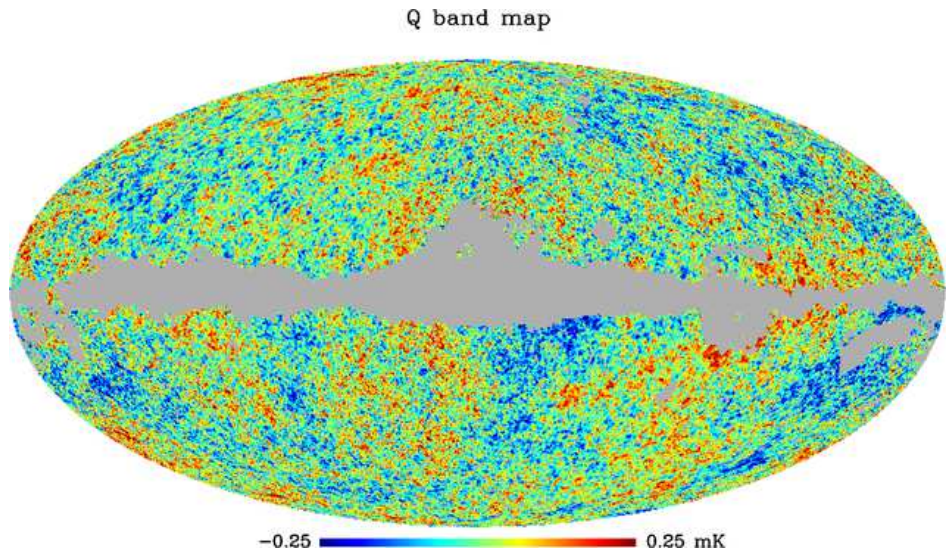

FIG. 1. *CMB radiation from WMAP data.*

where $R(\nu; x)$ denotes the emission at frequency $v$ for the corresponding temperature $T(x)$ (measured in Kelvin degrees), $c$ is the speed of light in the vacuum ($= 2.99798 \times 10^8$ m/s), $h$ is Planck's constant ($= 6.6261 \times 10^{-27}$ er g/s), and $k_B$ is Boltzmann's constant ($= 1.3807 \times 10^{-16}$ er g/K). In other words, the determination of $T(x)$ is made possible by the inversion of (1): the blackbody pattern can be estimated due to the presence of multiple detectors and the fact that astrophysical emissions of noncosmological nature are characterized by a different pattern of dependence across frequencies. In some regions, however, foreground emissions are so strong that component separation is still a difficult statistical problem; several groups of cosmologists are active in this field and a unique consensus solution has not been delivered yet. Moreover, in some areas of the sky (e.g., the Galactic plane, i.e., the line of sight of the Milky Way) the problem is considered to be largely unsolvable, so that there are missing observations in CMB maps (these unobserved regions are becoming, however, smaller and smaller with more refined experiments). In Figure 1 we report a CMB map constructed from (the Q band of) *WMAP* data; the missing region around the galactic plane is immediately evident.

Full-sky maps can be constructed by weighted linear interpolation across different channels, but they are not considered fully reliable for data analysis, especially at high frequencies; we report this so-called ILC (Internal Linear Combination) map in Figure 2, see Bennett et al. (2003) for more details on its construction.

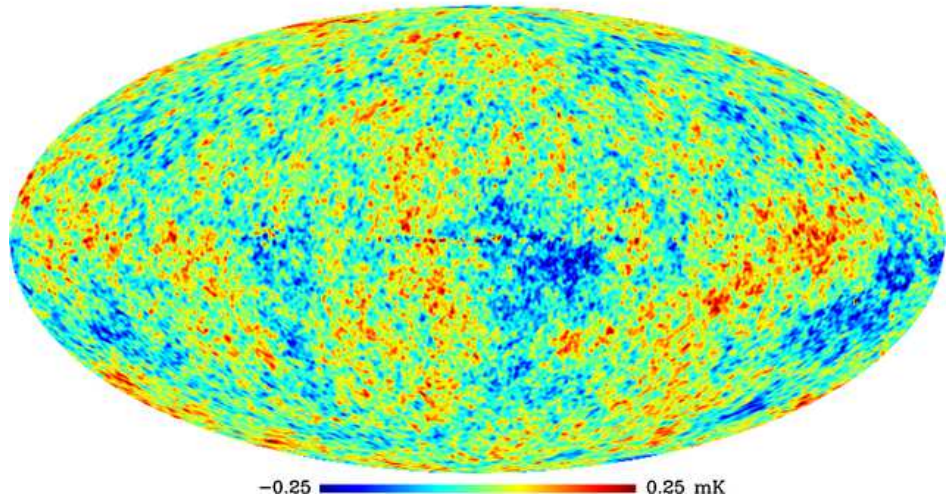

Fig. 2. *The so-called Internal Linear Combination map from WMAP data.*

There are several other statistically interesting issues involved with the reconstruction of the scalar value $T(x)$ from the vector-valued observations $\{O_1(x), \ldots, O_p(x)\}$; actually the real experimental set-up is more complicated (and interesting) than this, because each location is observed unevenly, that is, the *scanning strategy* is such that some regions are more accurately measured than others. Also, the contaminating noise can have a time-dependent structure [there is indeed strong evidence for long memory behavior, see, e.g., Natoli et al. (2002)]; the possible existence of noise correlation across different channels will be discussed below. These experimental features have sparked in the cosmological literature a very lively statistical debate on filtering and image reconstruction. We shall come back to some of these points later.

2.2. *Isotropy and spectral representation.* In the idealistic case of no experimental noise and perfect map-making, we can focus on the random field $\{T(x)\}$, assuming that it is exactly observed at each location on the unit sphere $S^2$. A crucial assumption on CMB radiation is its isotropic nature, that is, $T(\cdot) \stackrel{d}{=} T \circ g(\cdot)$, where $\stackrel{d}{=}$ denotes equality in distribution (in the sense of random fields) and $g \in SO(3)$ is any element of the group of rotations in $R^3$. More explicitly, the joint law of CMB radiation is assumed invariant to any change of coordinate; the condition is viewed by the physicists as a realization of so-called *Einstein's Cosmological Principle*, that is, the statement that the Universe should "look the same" to an observer in any arbitrary location. In other words, we could impose isotropy by requiring that the stochastic laws of CMB radiation are invariant with respect to the choice of coordinates. There is some (quite inconclusive) evidence from *WMAP* data that isotropy may fail, that is, some authors have suggested that data on CMB radiation may show some asymmetries which would be inconsistent with isotropy [see, e.g., Park (2004), Hansen et al. (2004)]. The existence of these asymmetries remains highly disputed, though, and it actually provides yet another intriguing area for statistical research. It is in fact hotly debated whether these asymmetries should be ascribed to experimental features or truly cosmological causes. From the theoretical point of view, cosmological models that would produce asymmetries do indeed exist, but they are highly nonstandard, ranging from global rotating solutions of Einstein's field equations to unconventional topological structures for the whole Universe. Much more methodological and applied research is needed in this area, but the question will most probably remain unsolved at least until the first releases of *Planck* data are available in a few years' time. By now, it is fair to say that a vast majority of cosmologists is still sticking to the isotropy assumption, and this is what we shall do in the present paper. Some of the procedures we shall consider in Section 4 for testing non-Gaussianity, however, are known to have also power against nonisotropic behavior; see, for instance, the local curvature approach below.

We shall hence focus on the statistical analysis of isotropic random fields. Throughout this paper we shall assume that the CMB random field is mean-square continuous, as it is always done in the CMB literature. Under the previous assumptions, the following spectral representation holds, in the mean square sense

$$T(x) = \sum_{l=0}^{\infty} \sum_{m=-l}^{l} a_{lm} Y_{lm}(x) \tag{2}$$

$$\text{where } a_{lm} = \int_{S^2} T(x) \overline{Y}_{lm}(x) \, dx. \tag{3}$$

Here, the bar denotes complex conjugation and $\{Y_{lm}(\cdot)\}$ the spherical harmonics, which form an orthonormal system for $L^2$ functions on the sphere. Some explicit expressions for the spherical harmonics can be found in the Appendix: much more complete treatment can be found elsewhere; see Varshalovich, Moskalev and Khersonskii (1988). For $l = m = 0$, we have $a_{00} = \int_{S^2} T(x) \, dx$, that is, the first coefficient is $4\pi$ times the sample mean of the random field. This value can be subtracted from $T(x)$, whence we can take the expansion to start from $l = 1$; indeed, in practice, in the cosmological literature also the coefficients corresponding to $l = 1$ are discarded (the so-called dipole terms), as they have no cosmological meaning, but they simply reflect the absolute motion of the Earth with respect to the frame of reference with respect to which CMB radiation is at rest. For $l \geq 2$, the triangular array $\{a_{lm}(\cdot)\}$ represents zero-mean, complex-valued random coefficients, with variance $E|a_{lm}|^2 = C_l > 0$, the angular power spectrum of the random field. The coefficients are uncorrelated, $Ea_{l_1 m_1} \overline{a}_{l_2 m_2} = C_{l_1} \delta_{l_1}^{l_2} \delta_{m_1}^{m_2}$, and, hence, in the Gaussian case they are independent [note, however, that $a_{lm} = (-1)^m \overline{a}_{l-m}$]. We have the identity

$$E\left\{\sum_{l=2}^{\infty} \sum_{m=-l}^{l} a_{lm} Y_{lm}(x)\right\}^2 = \sum_{l=2}^{\infty} \sum_{m=-l}^{l} E|a_{lm}|^2 Y_{lm}(x)$$
$$= \sum_{l=2}^{\infty} C_l \sum_{m=-l}^{l} Y_{lm}(x) = \sum_{l=2}^{\infty} C_l \frac{2l+1}{4\pi},$$

in view of a standard summation formula for spherical harmonics [Varshalovich, Moskalev and Khersonskii (1988)]. It follows immediately that $C_l(2l + 1)$ must be summable to ensure finite variance. The angular power spectrum in the Gaussian case provides a complete characterization of the dependence structure of the random field; to its estimation from CMB data we now turn our attention.

## 3. Angular power spectrum estimation.

3.1. *Power spectrum estimation under idealistic circumstances.* As noted before, having observed the random field $T(x)$, the coefficients $\{a_{lm}(\cdot)\}$ can be recovered by means of the inverse Fourier transform (3). In practice, with real data the integral is replaced by finite sums by means of (exact or approximate) cubature formulae, which are implemented in standard packages for CMB data analysis such as HealPix or GLESP [see Gorski et al. (2005), Doroshkevich et al. (2005)]. The angular power spectrum can then be estimated by

$$\widehat{C}_l = \frac{1}{2l+1} \sum_{m=-l}^{l} |a_{lm}|^2. \tag{4}$$

This simple estimator highlights a very important issue when dealing with CMB data. It is indeed readily seen that the estimator is consistent in the Gaussian case, as $l \to \infty$; more precisely,

$$E\widehat{C}_l = C_l,$$
$$E\left\{\frac{\widehat{C}_l}{C_l} - 1\right\}^2 = \frac{1}{(2l+1)^2} E\left[\frac{a_{l0}^2}{C_l} - 1 + 2\left\{\sum_{m=1}^{l} \frac{|a_{lm}|^2}{C_l} - 1\right\}\right]^2$$
$$= \frac{2}{2l+1} = o(1),$$

because $a_{l0}^2/C_l \stackrel{d}{\sim} \chi_1^2$ and for $m=1,\ldots,l$, $2a_{lm}^2/C_l \stackrel{d}{\sim}$ i.i.d. $\chi_2^2$, where $\chi_n^2$ denotes a standard chi-square random variable with $n$ degrees of freedom. In the Gaussian case with fully observed maps, the issue of angular power spectrum estimation can thus be considered trivial, and indeed, the previous expressions not only ensure consistency but they also provide exact confidence intervals: it is immediate to see that

$$\sum_{m=-l}^{l} |a_{lm}|^2 = \left\{ |a_{l0}|^2 + \sum_{m=-1}^{l} 2|a_{lm}|^2 \right\} \stackrel{d}{\sim} C_l \times \chi_{2n+1}^2.$$

However, we must stress that these results rely heavily on the Gaussian assumption. Indeed, Baldi and Marinucci (2007) and Baldi, Marinucci and Varadarajan (2007) have shown that under isotropy the coefficients $a_{lm}$ can only be independent in the Gaussian case, despite the fact that they are always uncorrelated by construction: in other words, sampling independent, non-Gaussian random coefficients to generate maps according to (2) will always yield an anisotropic random field. The correlation structure of the coefficients $\{a_{lm}\}$ is in general quite complicated, despite the fact that it can be very nicely characterized in terms of group representation properties for $SO(3)$ [Marinucci and Peccati (2007)]. In view of this, to derive any asymptotic result for $\widehat{C}_l$ under non-Gaussianity is by no means trivial; indeed, even the possible consistency (as $l \to \infty$) of the estimator (4) in non-Gaussian circumstances is still an open issue for research.

3.2. *Dealing with instrumental noise.* We shall now try to make our analysis more realistic by considering the effect of noise and missing observations. Starting from the former, we shall consider the case where we observe $O(x) := T(x) + N(x)$, $N(x)$ denoting instrumental noise; for simplicity, we shall follow the cosmological literature, assuming $N(x)$ to be also a zero mean, mean square continuous and isotropic random field on the sphere. Whereas the assumptions of zero-mean and mean square continuity are basically immaterial, isotropy of the noise may need to be relaxed if the sky is unevenly observed. We shall also assume that $T(x)$ and $N(x)$ are independent. Performing the spherical harmonic transform, we obtain, in an obvious notation,

$$a_{lm} = \int_{S^2} \{T(x) + N(x)\}\overline{Y}_{lm}(x)\,dx =: a_{lm}^T + a_{lm}^N,$$

which leads to

$$\widehat{C}_l = \frac{1}{2l+1}\left[\sum_{m=-l}^{l} |a_{lm}^T|^2 + \sum_{m=-l}^{l} |a_{lm}^N|^2 + 2\operatorname{Re}\left\{\sum_{m=-l}^{l} a_{lm}^S \overline{a}_{lm}^N\right\}\right].$$

It is immediate to see that the resulting estimator is biased, $E\widehat{C}_l = C_l^T + C_l^N$; the variance is easily seen to be given by

$$\text{Var}\{\widehat{C}_l\} = \frac{2\{C_l^T + C_l^N\}}{2l+1}. \tag{5}$$

In the cosmological literature, the standard procedure to address this bias is to assume that the noise correlation structure can be derived by Monte Carlo simulations or instrumental calibration; under this assumption, it is possible to subtract the bias from $\widehat{C}_l$ and obtain a correct estimator with variance (5). An obvious question is then to test whether the assumption that $C_l^N$ is known does not introduce some spurious effect into the analysis (namely, some unaccounted bias). A proposal in this direction was put forward by Polenta et al. (2005). To understand this idea, we must get back to the multi-channel setting, where we observe

$$O_i(x) := T(x) + N_i(x), \qquad i = 1, \ldots, p,$$

which in the harmonic domain leads to

$$a_{i;lm} := a_{lm}^T + a_{lm}^{N_i}.$$

Note that the temperature component of the random spherical harmonics coefficients does not depend on the observing channel. We assume that the noise is independent over channels, which is believed to be consistent with the actual experimental set-ups of current datasets. Testing noise correlation across different channels is yet another open challenge for research. For a given noise structure, an obvious estimator for $C_l$ is

$$\widetilde{C}_l^A := \frac{1}{p} \sum_{i=1}^{p} \{\widehat{C}_{il} - C_l^{N_i}\},$$

$$\widehat{C}_{il} := \frac{1}{2l+1} \sum_{m=-l}^{l} |a_{i;lm}|^2. \tag{6}$$

The estimator $\widetilde{C}_l^A$ is known in the literature as the *auto-power spectrum*. Simple computations yield [Polenta et al. (2005)]

$$E\widetilde{C}_l^A = C_l,$$

$$\text{Var}\{\widetilde{C}_l^A\} = \frac{2}{2l+1} \left\{ C_l^2 + \frac{2C_l}{p^2} \sum_{i=1}^{p} C_l^{N_i} + \frac{1}{p^4} \sum_{i,j=1}^{p} C_l^{N_i} C_l^{N_j} \right\}.$$

Of course, the natural question that arises at this stage is the possible existence of misspecification, that is, some errors in the bias-correction term $C_l^{N_i}$. A solution for this issue was proposed by Polenta et al. (2005). The idea is to focus on the *cross-power spectrum* estimator

$$\widetilde{C}_l^{CP} = \frac{2}{p(p-1)} \sum_{i=1}^{p-1} \sum_{j=i+1}^{p} \left( \frac{1}{2l+1} \sum_{m=-l}^{l} a_{i;lm} \overline{a}_{j;lm} \right).$$

The underlying rationale for $\widetilde{C}_l^{CP}$ is easy to gather: under the assumption that noise is independent across a different channel, the estimator is unbiased, regardless of the value of the $C_l^{N_i}$. More precisely,

$$E\widetilde{C}_l^{CP} = \frac{2}{p(p-1)} \sum_{i=1}^{p-1} \sum_{j=i+1}^{p} \left( \frac{1}{2l+1} \sum_{m=-l}^{l} E(a_{lm}^T + a_{lm}^{N_i})(\overline{a}_{lm}^T + \overline{a}_{lm}^{N_j}) \right)$$

$$= \frac{2}{p(p-1)} \sum_{i=1}^{p-1} \sum_{j=i+1}^{p} C_l = C_l.$$

Similar manipulations yield

$$\text{Var}\{\widetilde{C}_l^{CP}\} = \frac{2}{2l+1} \left\{ C_l^2 + \frac{2C_l}{p^2} \sum_{i=1}^{p} C_l^{N_i} + \frac{1}{p^2(p-1)^2} \sum_{i=1}^{p-1} \sum_{j=i+1}^{p} C_l^{N_i} C_l^{N_j} \right\}.$$

Merely for notational simplicity, we also assume that the noise variance is constant across detectors. It is then readily seen that

$$\text{Var}\{\widetilde{C}_l^{CP}\} - \text{Var}\{\widetilde{C}_l^A\} = \frac{2}{2l+1} \left\{ \frac{1}{p^2(p-1)} (C_l^N)^2 \right\}.$$

More explicitly, the auto-power spectrum estimator is more efficient that the cross-power spectrum; however, the latter is robust to noise misspecification. This is the classical setting which makes the implementation of a Hausman-type test for misspecification feasible [Hausman (1978)]. Indeed, it is possible to consider the statistic

$$H_l = [\text{Var}\{\widetilde{C}_l^{CP} - \widetilde{C}_l^A\}]^{-1/2} \{\widetilde{C}_l^{CP} - \widetilde{C}_l^A\},$$

$$\text{Var}\{\widetilde{C}_l^{CP} - \widetilde{C}_l^A\} = \frac{2}{2l+1} \left\{ \frac{1}{p^4} \sum_{i=1}^{p} \{C_l^{N_i}\}^2 + \frac{2}{(p-1)^2} \sum_{i=1}^{p-1} \sum_{j=i+1}^{p} C_l^{N_i} C_l^{N_j} \right\}.$$

Under the null of exact bias correction, it is readily seen that $H_l \to_d N(0,1)$, as $l \to \infty$. On the other hand, in the presence of misspecification, that is, when the actual noise variance is equal to $C_l^{N_i} + \delta$ for some $i$, $\delta > 0$, then we expect $EH_l$ to diverge with rate $\sqrt{l}\delta$ as $l \to \infty$.

It is also possible to consider a functional form of the same test, focusing on

$$B_L(r) := \frac{1}{\sqrt{L}} \sum_{l=1}^{[Lr]} H_l, \qquad r \in [0,1].$$

It is standard to show that $B_L(r)$ converges weakly to a standard Brownian motion, as $L \to \infty$. A test for noise misspecification can then be constructed along the lines of standard Kolmogorov–Smirnov or Cramér–Von Mises statistics. We refer again to Polenta et al. (2005) for a much more detailed discussion and an extensive simulation study.

The methods discussed above rely on a basic identification assumption, that is, the condition that instrumental noise be independent across different channels. This is an assumption which is commonly entertained in the cosmological literature; suitable statistical issues to test its validity are still lacking and represent an open issue for research. A more challenging research task was mentioned before: the previous discussion was entirely led under the assumption that the CMB field (and thus the corresponding spherical harmonics coefficients) are Gaussian. It is very important to stress that relaxing this assumption has much deeper consequences here than it is usually the case in statistical inference. Indeed, it follows from results in Baldi and Marinucci (2007) that *if the field is isotropic, the coefficients $(a_{lm})$ cannot be independent unless they are Gaussian.* It follows that even the simple consistency (as $l \to \infty$) of the estimator $\widetilde{C}_l$ remains an open issue to address, in general non-Gaussian circumstances. We shall not go further into this issue here, but we rather focus on another important feature of realistic datasets: the presence of unobserved regions, which make the exact evaluation of the inverse Fourier transform (3) unfeasible.

3.3. *Missing observations.* The presence of missing observations, that is, regions of the sky where the CMB is deeply contaminated by astrophysical foregrounds, posits serious challenges to angular power spectrum estimation. The first consequence is that the sample spherical harmonics coefficients

$$a_{lm}^M = \int_{S^2/M} T(x)\overline{Y}_{lm}(x)\,dx,$$

lose their uncorrelation properties (here, $M$ denotes the unobserved region and, for notational simplicity, we came back to the case of a single detector with no instrumental noise). Indeed, we have

$$
\begin{aligned}
Ea_{l_1m_1}^M \overline{a}_{l_2m_2}^M &= E\left\{\left(\int_{S^2/M} T(x)\overline{Y}_{l_1m_1}(x)\,dx\right)\left(\int_{S^2/M} T(y)Y_{l_2m_2}(y)\,dy\right)\right\} \\
&= \sum_{l_1m_1}\sum_{l_2m_2} Ea_{lm}\overline{a}_{l'm'}\left(\int_{S^2/M} Y_{lm}(x)\overline{Y}_{l_1m_1}(x)\,dx\right) \\
&\qquad \times \left(\int_{S^2/M} \overline{Y}_{l'm'}(y)Y_{l_2m_2}(y)\,dy\right) \\
&= \sum_{lm} C_l W_{lml_1m_1} W_{lml_2m_2},
\end{aligned}
\tag{7}
$$

where

$$W_{lml_1m_1} := \int_{S^2/M} Y_{lm}(x)\overline{Y}_{l_1m_1}(x)\,dx.$$

In case the spherical random field is fully observed, then $M = \varnothing$ (the empty set) and by standard orthonormality properties of the spherical harmonics $Y_{lm}$, we obtain $W_{lml_1m_1} = \delta_{l_1}^l \delta_{m_1}^m$ and, therefore, $Ea_{l_1m_1}\overline{a}_{l_2m_2} = C_l \delta_{l_1}^{l_2} \delta_{m_1}^{m_2}$. In the presence of missing observations, the random coefficients are no longer uncorrelated neither over $l$ nor over $m$. In the physical literature the values of $\{W_{lml_1m_1}\}_{l_1m_1l_2m_2}$ are computed numerically, exploiting the a priori knowledge on the geometry of the unobserved regions; the resulting *coupling matrices* can then be used to deconvolve the estimated values $\widehat{C}_l$, a procedure which has become extremely popular under the name of *MASTER* [see Hivon et al. (2002) for details]. In practice, it is not possible to identify by this method the value of the angular power spectrum at every single multipole $l$; it is then customary to proceed with *binning techniques*, where the values of $C_l$ at nearby frequencies are averaged and only these smoothed values are actually estimated. Plots for the estimates of the $C_l$ derived along these lines can be found, for instance, on the web site of *WMAP*; a comparison with angular power spectrum estimate from several other experiments (based upon smaller patches of the observed sky) is also entertained.

The previous procedures can be computationally extremely demanding and we would like here to introduce an alternative strategy, which was basically put forward in Baldi et al. (2006). The idea is to implement power spectrum estimation by means of new kinds of spherical wavelets, called needlets [see also Narcowich, Petrushev and Ward (2006a), Narcowich, Petrushev and Ward (2006b), Marinucci et al. (2008) and Baldi et al. (2007)]. Needlets can be described as a convolution of the spherical harmonics basis by means of a suitable kernel function $b(\cdot)$; more precisely, the general element of the needlet frame can be written down as

$$\psi_{jk}(x) = \sqrt{\lambda_{jk}} \sum_{l=B^{j-1}}^{B^{j+1}} b\left(\frac{l}{B^j}\right) \sum_{m=-l}^{l} Y_{lm}(x)\overline{Y}_{lm}(\xi_{jk}),$$

where $\{\xi_{jk}\}$ denotes a set of grid points on the sphere, $B > 1$ is a bandwidth parameter, $b(\cdot)$ is compactly supported and an infinitely differentiable function which satisfies the *partition-of-unity* property, that is,

(8) $$\sum_j b^2\left(\frac{l}{B^j}\right) \equiv 1 \qquad \text{for all } l > 1,$$

and $\{\lambda_{jk}, \xi_{jk}\}$ (the *cubature points* and *cubature weights*) can be chosen in such a way that

$$\sum_k Y_{l_1m_1}(\xi_{jk})\overline{Y}_{l_2m_2}(\xi_{jk})\lambda_{jk} = \int_{S^2} Y_{l_1m_1}(x)\overline{Y}_{l_2m_2}(x)\,dx = \delta_{l_1}^{l_2}\delta_{m_1}^{m_2}.$$

More details on this construction and its underlying rationale can be found in Baldi et al. (2006) and are not reported here for brevity's sake; see also Kerkyacharian et al. (2007) and Guilloux, Fay and Cardoso (2007) for further work in this area. The corresponding random needlets coefficients are provided by the analysis formula

$$\widehat{\beta}_{jk} = \int T(x)\psi_{jk}(x)\,dx = \sqrt{\lambda_{jk}} \sum_{l=B^{j-1}}^{B^{j+1}} \sum_{m=-l}^{l} b\left(\frac{l}{B^j}\right) a_{lm} \overline{Y}_{lm}(\xi_{jk}),$$

whereas the synthesis expression is given as

$$\begin{aligned}
\sum_{j,k} \beta_{jk}\psi_{jk}(x) &= \sum_{j} \sum_{l_1=B^{j-1}}^{B^{j+1}} \sum_{m_1=-l_1}^{l_1} b\left(\frac{l_1}{B^j}\right) b\left(\frac{l_2}{B^j}\right) a_{l_1 m_1} Y_{l_1 m_1}(x) \\
&\quad \times \sum_{l_2=B^{j-1}}^{B^{j+1}} \sum_{m_2=-l_2}^{l_2} \sum_{k} Y_{l_1 m_1}(\xi_{jk}) \overline{Y}_{l_2 m_2}(\xi_{jk}) \lambda_{jk} \\
&= \sum_{j} \sum_{l_1=B^{j-1}}^{B^{j+1}} \sum_{m_1=-l_1}^{l_1} b\left(\frac{l_1}{B^j}\right) b\left(\frac{l_2}{B^j}\right) a_{l_1 m_1} Y_{l_1 m_1}(x) \\
&\quad \times \sum_{l_2=B^{j-1}}^{B^{j+1}} \sum_{m_2=-l_2}^{l_2} \delta_{l_1}^{l_2} \delta_{m_1}^{m_2} \\
&= \sum_{l=1}^{\infty} \sum_{m=-l}^{l} a_{lm} Y_{lm}(x) = T(x),
\end{aligned}$$

using (8). For our purposes, it is sufficient to recall the main properties of the needlets construction:

- needlets enjoy excellent localization properties in the real domain, each $\psi_{jk}(x)$ being quasi-exponentially localized around its center $\xi_{jk}$. As such, needlets coefficients have been shown to be minimally influenced by the presence of missing observations.
- the needlets system is compactly supported in the harmonic domain; as such, the random needlets coefficients are uncorrelated for $j - j' \geq 2$. Much more surprisingly, the random needlets coefficients are asymptotically uncorrelated for any fixed angular distance, as the frequency $j$ diverges to infinity. This property implies that (in the Gaussian case) it is possible to derive a growing array of asymptotically i.i.d. observations out of a single realization of an isotropic random field. This opens the way to a plethora of statistical procedures.

In particular, it is possible to suggest the estimator

$$\widehat{\Gamma}_j := \sum_k \widehat{\beta}_{jk}^2 = \sum_k \left\{ \sum_{l=B^{j-1}}^{B^{j+1}} \sum_{m=-l}^{l} \sqrt{\lambda_{jk}} b\left(\frac{l}{B^j}\right) a_{lm} \overline{Y}_{lm}(\xi_{jk}) \right\}^2$$

$$= \sum_{l_1,l_2=B^{j-1}}^{B^{j+1}} \sum_{m_1,m_2} b\left(\frac{l_1}{B^j}\right) b\left(\frac{l_2}{B^j}\right) a_{l_1 m_1} \overline{a}_{l_2 m_2} \left\{ \lambda_{jk} \sum_k Y_{lm}(\xi_{jk}) \overline{Y}_{lm}(\xi_{jk}) \right\}$$

$$= \sum_{l=B^{j-1}}^{B^{j+1}} b^2\left(\frac{l}{B^j}\right) \widehat{C}_l (2l+1),$$

for which it is simple to show that

$$(9) \qquad E\widehat{\Gamma}_j = \sum_{l=B^{j-1}}^{B^{j+1}} b^2\left(\frac{l}{B^j}\right) C_l (2l+1).$$

Equation (9) shows that $\widehat{\Gamma}_j$ provides an unbiased estimator for a smoothed version of the angular power spectrum; the advantage with respect to the standard procedure is that not only unbiasedness, but even uncorrelation over different scales is asymptotically conserved in the presence of missing observations, making the implementation of confidence intervals and testing procedures viable [see again Baldi et al. (2006) for details]. Also, even in the presence of a masked region, the summands $\{\widehat{\beta}_{jk}\}$ are still asymptotically independent (over $k$) as $j \to \infty$, whereas we have seen in (7) that this is not the case for the random coefficients $\{a_{lm}\}$. The price for such robustness properties is clearly connected to the smoothing, that is, in the presence of missing observations it turns out to be unfeasible to estimate each angular power spectrum mode $C_l$ by itself, and one must stick to a slightly less ambitious goal, that is, the estimation of joint values averaged over some subset of frequencies (chosen by the data analyst). There is, of course, a standard trade-off in the choice of the bandwidth parameter $B$: values closer to unity entail a much better resolution, but this brings about worse localization properties on the sphere and therefore a possibly higher contamination from spurious observations; on the other hand, higher values of $B$ yield more robust, but less informative estimates.

Spherical wavelets in general, and needlets in particular, allow for many statistical applications, which go much beyond angular power spectrum estimation. One example is the analysis of cross-correlation between CMB and Large Scale Structure (LSS) maps; this is a key prediction of many cosmological models entailing some form of *dark energy* and has been implemented on real data by Pietrobon, Balbi and Marinucci (2006). Other applications may include testing for non-Gaussianity and isotropy, bootstrap/subsampling evaluation of confidence intervals for CMB statistics [Baldi et al. (2007)], component separation and many others. Given such a wide array of applications, we stress the need for a more careful analysis of their theoretical underpinnings, with special reference to the effect of the Gaussianity assumption on our conclusions. This and many other related issues are left as topics for further research.

3.4. *Parameter estimation.* In this paper we shall neglect almost completely another crucial issue in CMB data analysis, which is very tightly coupled to the estimation of the angular power spectrum, that is, cosmological parameter estimation. More precisely, the theoretical angular power spectrum can be written as a function of a number of cosmological parameters, such as the *baryon, matter and dark energy densities* $\Omega_b, \Omega_m, \Omega_\Lambda$, the *optical depth* $\tau$, the *spectral index* $n_s$, the *Hubble constant* $H_0$ and others; of course, the numbers of parameters to be estimated varies across different cosmological models, typically ranging from 6 to 16; see again Dodelson (2003) for more details. There are no known closed-form expressions yielding the theoretical angular power spectrum $C_l$ as an explicit function of these parameters (which we write for brevity as $\vartheta$); however, there are indeed very fast numerical routines which solve the associated partial differential equations and provide as an output $C_l$ after a specific value of $\vartheta$ has been supplied [see Seljak and Zaldarriaga (1996)].

Once the set of estimated values $\widehat{C}_l$ has also been derived, there are basically two approaches that have been implemented to obtain estimates for the set of parameters, namely, some form of minimum distance estimators, where the parameters are calibrated to minimize a weighted distance between $C_l(\vartheta)$ and $\widehat{C}_l$, and approximate maximum likelihood methods, where suitable approximations for the likelihood functions are derived and the estimates are consequently derived. In practice, both methods are implemented by means of a heavy use of numerical techniques (especially MCMC), and a lively debate is growing on the construction of the most efficient algorithms. Likewise, an extensive discussion is growing on the construction of confidence intervals for the parameters, where fundamental issues such as the differences between Bayesian and frequentist viewpoints are often called upon (the distinction between these two approaches is not perceived in the cosmological community in the same manner as in the statistical one; just to give an example, maximum likelihood estimates are nearly unanimously labeled a Bayesian procedure in the CMB literature).

For brevity's sake, we are unable to go deeper into these issues, which are still quite far from a satisfactory solution. We refer, for instance, to Hamann and Wong (2008) and the references therein for more discussion and recent proposals in this area.

**4. Testing for non-Gaussianity.** Among the several statistical issues which arise in connection to the analysis of Cosmic Microwave Background radiation, a lot of attention has been drawn by non-Gaussianity tests. These tests have several motivations. The first is connected to the need for a statistical validation for the predictions of the so-called inflationary scenario, which is currently the leading incumbent as a standard model for the Big Bang dynamics; see Dodelson (2003) for discussions and explanations. Under this labeling, there exist an enormous variety of different physical models, which in a vast majority of circumstances lead to expressions such as

$$(10) \qquad T(x) = T_G(x) + f_{NL}\{T_G^2(x) - ET_G^2(x)\},$$

where $T(x)$ denotes as before CMB, $T_G(x)$ is an underlying Gaussian field, $f_{NL}$ is a nonlinearity parameter and the unit of measurements are such that the non-Gaussian part $T_G^2(x) - ET_G^2(x)$ is $10^{-4}/10^{-5}$ times smaller than $T_G(x)$. (10) should be viewed as a strong simplification, for several reasons: in particular, we are considering exclusively the primordial dynamics, thus neglecting later interactions through the gravitational potential; also, we are ruling out more complicated models, where higher order terms or multiple subordinating fields may be present; and, of course, we are neglecting a whole plethora of observational issues, where possible non-Gaussianities may be formed by secondary effects, such as the interactions of incoming photons at more recent epochs. Despite all these simplifying conditions, (10) does provide an extremely good guidance for features to be expected and, indeed, it makes up a benchmark model against which many procedures have been tested in the last few years. In particular, considerable attention has been drawn by the possibility to constrain the value of $f_{NL}$, as this depends on constants from fundamental physics [Bartolo et al. (2004)] and as such it allows to probe many features of cosmological models.

Among several statistical procedures which have been proposed in the literature, we shall focus on three main families, namely, tests based upon the bispectrum, tests based upon geometric features of Gaussian random fields (local curvature) and tests based upon spherical wavelets (in this case, so-called Spherical Mexican Hat Wavelets).

4.1. *The angular bispectrum.* It is obvious that, under Gaussianity, the sequence $\{a_{lm}\}$, $m = 0, \ldots, l$ makes up an array of independent Gaussian random variables (complex-valued for $m \neq 0$), so that a natural first option for a test of Gaussianity is to consider their sample skewness $a_{l_1m_1}a_{l_2m_2}a_{l_3m_3}$ and check whether it is significantly different from zero. This simple idea is made much more sophisticated by the necessity to impose rotational invariance on the sample coefficients. Such invariance can be imposed by demanding that the probability law of the CMB field be invariant with respect to the action of the rotation group. More formally, let $g \in SO(3)$ be any element of the rotation group in $\mathbb{R}^3$; the assumption of isotropy can then be written as

$$T(x) \stackrel{d}{=} T(gx) \qquad \text{for all } x \in S^2,$$

whereas in terms of the spectral representation, we have

(11) $$\sum_{l=0}^{\infty} \sum_{m=-l}^{l} a_{lm} Y_{lm}(x) \stackrel{d}{=} \sum_{l=0}^{\infty} \sum_{m=-l}^{l} a_{lm} Y_{lm}(gx).$$

As explained, for instance, in Hu (2001) and Marinucci and Peccati (2007), from (11) it follows that the bispectrum of a rotational invariant random field must take the form

$$E a_{l_1 m_1} a_{l_2 m_2} a_{l_3 m_3} = \begin{pmatrix} l_1 & l_2 & l_3 \\ 0 & 0 & 0 \end{pmatrix} \begin{pmatrix} l_1 & l_2 & l_3 \\ m_1 & m_2 & m_3 \end{pmatrix} b_{l_1 l_2 l_3},$$

where $b_{l_1 l_2 l_3}$ (the reduced bispectrum) conveys the physical information and does not depend on $m_1, m_2, m_3$. The Wigner's $3j$ symbols appearing on the right-hand side are discussed in the Appendix; many more details can be found, for instance, in Varshalovich, Moskalev and Khersonskii (1988), Marinucci (2006) and Marinucci (2008), whereas generalization to higher order cumulant spectra are described in Marinucci and Peccati (2008). A feasible, rotationally invariant estimator for the (normalized) bispectrum is provided by

$$I_{l_1 l_2 l_3} = (-1)^{(l_1+l_2+l_3)/2} \sum_{m_1 m_2 m_3} \begin{pmatrix} l_1 & l_2 & l_3 \\ m_1 & m_2 & m_3 \end{pmatrix} \frac{a_{l_1 m_1} a_{l_2 m_2} a_{l_3 m_3}}{\sqrt{C_{l_1} C_{l_2} C_{l_3}}}$$

and its studentized version is of course

$$\widehat{I}_{l_1 l_2 l_3} = (-1)^{(l_1+l_2+l_3)/2} \sum_{m_1 m_2 m_3} \begin{pmatrix} l_1 & l_2 & l_3 \\ m_1 & m_2 & m_3 \end{pmatrix} \frac{a_{l_1 m_1} a_{l_2 m_2} a_{l_3 m_3}}{\sqrt{\widehat{C}_{l_1} \widehat{C}_{l_2} \widehat{C}_{l_3}}}.$$

The sample bispectrum is discussed, for instance, by Hu (2001); asymptotic properties are provided by Marinucci (2006) and Marinucci (2008), where the phase factor $(-1)^{(l_1+l_2+l_3)/2}$ is also introduced. In particular, it can be shown that the sequences $\{I_{l_1 l_2 l_3}\}$ and $\{\widehat{I}_{l_1 l_2 l_3}\}$ converge to Gaussian independent random variables in the high frequency limit where $\min(l_1, l_2, l_3) \uparrow \infty$. The limiting behavior of the bispectrum ordinates, however, is perhaps not the most significant instrument for the implementation of statistical procedures. More precisely, it seems more promising to combine the different ordinates into a single statistic, by means of the integrated bispectra

$$J_{1L}(r) = \sum_{l_2=1}^{[Lr]} \left\{ \frac{1}{\sqrt{K}} \sum_{k=1}^{K} \widehat{I}_{l_1+k, l_2 l_2} \right\}, \qquad J_{2L}(r) = \sum_{l=1}^{[Lr]} \widehat{I}_{lll}.$$

Convergence to Brownian motion for both these statistics was established in Marinucci (2006). The underlying rationale can be briefly motivated as follows: in both cases we combine several different ordinates into a single functional statistic, capable of keeping track of the frequency location for possible deviations from Gaussianity. The different combination of multipoles in $J_{1L}(\cdot), J_{2L}(\cdot)$ corresponds to the two well-known classes of squeezed and equilateral configurations, as discussed again by Marinucci (2006), Babich, Creminelli and Zaldarriaga (2004) and many others. It is also possible to provide some results on the asymptotic behavior of these statistics under non-Gaussian circumstances; in particular, results in Marinucci (2006) suggest that $J_{1L}$ will provide consistent testing procedures (as $L \to \infty$) under model (10), whereas tests based upon $J_{2L}$ will have asymptotically negligible power, for all values of $f_{NL}$. These theoretical findings have been validated by Monte Carlo simulations in Cabella et al. (2006); the integrated bispectrum has also been shown to compare favorably with alternative statistical procedures in some internal statistical challenges within the Planck collaboration.

In Figure 3 we report the results obtained by implementing $J_{1L}(r)$ on the data from the (2003) and (2007) WMAP data releases. We stress that the simulations are calibrated in a realistic experimental setting, that is, they do take into account features such as the presence of noise and missing observations. More precisely, we used 1000 simulated maps of CMB signal plus noise; we took into account the modulation of noise on the maps given by *WMAP* scanning strategy, the presence of a masked region to avoid the emission from the Milky Way and point sources, and we considered the optical transfer function of the telescopes. To comply with the cosmological literature, the shaded region represents the 68% confidence interval ($1\sigma$) as evaluated by means of Monte Carlo simulations for various values of $r \in [0,1]$: we fixed $L = 500$ because *WMAP* data allow a reliable coverage up to this multipoles; see Bennett et al. (2003) and Hinshaw et al. (2008) for more technical details on the experiment (it should be noted that in the $x$-axis we report $rL$). The dotted and dashed lines represent Monte Carlo expected values for our statistics with $f_{NL} = \pm 100, \ldots, 500$, respectively. It is possible to check that the boundary value of $f_{NL}$ to ensure detection is in the order of 200 or larger, that is, with a signal to noise ratio in the order of a few percentage points. This is indeed confirmed by a more detailed study in Cabella et al. (2006). Finally, triangles (2003 dataset) and squares (2007 dataset) represent the evaluation of the statistic on real data, on the basis of the previously mentioned *WMAP* releases. It is clear that the evidence for non-Gaussianity is rather weak, and, indeed, the statistics get closer to zero as the observations increase. We must stress, however, that the level of non-Gaussianity favored by theorists is well below 100, and this is still consistent with observations at the current resolution. Note that the signal to noise ratio for the non-Gaussian signal is in the order of $f_{NL}/10^4$, so that these values are really difficult to detect.

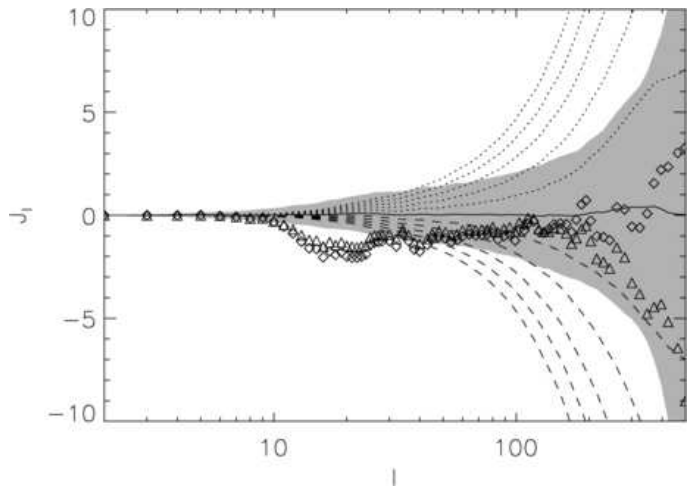

Fig. 3. *The behavior of $J_{1L}(r)$ on WMAP data.*

Very recently, in Yadav and Wandelt (2007) it has been claimed a detection of a nonzero $f_{NL}$ ($\simeq 80$) by means of a modified bispectrum estimator, which is constructed to take into account the presence of noise and missing observations, at the same time keeping computational costs at a feasible level. This proposal is indeed very interesting; the results, however, are quite close to the boundary level and as such they must probably be considered not conclusive. The general consensus in the community seems to be that new releases of data from more sophisticated experiments such as *Planck*, and possibly more efficient statistical procedures yet to be devised, will indeed be necessary to settle the question on the possible existence of non-Gaussianity in CMB. It should be stressed, in particular, that the bispectrum requires the evaluation of the inverse Fourier transform (3), and as such it is known to be severely affected by the presence of missing observations (there is some evidence that the detection level could reach $f_{NL} \simeq 10$ or lower for fully observed sky maps). Improving the performance of the bispectrum for partial sky coverage is a priority of current research in view of the forthcoming satellite data: for instance, in Lan and Marinucci (2008) the bispectrum approach is combined with the needlets construction described in the previous section. Rather than considering these further developments in the bispectrum literature, we move to other methods which have a local nature, and are thus expected to be more robust in the presence of missing data.

4.2. *Local curvature.* The next approach to constraining possible non-Gaussianities is based upon an analysis of the local features of Gaussian random fields. This is an issue which has a long tradition in probability, as summarized, for instance, in the recent book by Adler and Taylor (2007). In this respect, many efforts have focused on excursion sets and other procedures from convex geometry (the so-called Minkowski functionals); these ideas have found many fruitful applications in CMB data analysis; see, for instance, Hikage et al. (2008) and the references therein.

However, our approach here will be different, and more directly rooted to differential geometry; we collect in the Appendix some background material to understand better the notation and the approach. The intuition can be explained as follows. At any point $x \in S^2$ on the sphere, it is possible to investigate the local curvature of the random field $T$ by focusing on its *covariant* Hessian matrix; in particular, we can study whether this Hessian defines a positive definite bilinear form (in which case we will label $x$ as *lake* point), a negative definite form (in which case we will label $x$ as *hill* point) or neither of the two (in which case we will label $x$ as *saddle* point). This approach was proposed by Dore', Colombi and Bouchet (2003), in the standard Euclidean circumstances, and then applied to the spherical case by Hansen et al. (2004) and Cabella et al. (2005). Here, a crucial step is to ensure that the Hessian is defined in such a way to have an intrinsic meaning, that is, the geometric characterization of the points must be independent from the choice of coordinates. The appropriate instrument for this point is the notion of *covariant derivative*, which we recall briefly in the Appendix. We are finally in the position to evaluate the covariant Hessian of any random field on the sphere, which is provided by

$$
\begin{aligned}
H :=& \begin{pmatrix} T_{;\vartheta\vartheta} & T_{;\vartheta\varphi}/\sin\vartheta \\ T_{;\vartheta\varphi}/\sin\vartheta & T_{;\varphi\varphi}/\sin^2\vartheta \end{pmatrix} \\
=& \begin{pmatrix} T_{,\vartheta\vartheta} & (T_{,\vartheta\varphi} - \cot\vartheta T_{,\varphi})/\sin\vartheta \\ (T_{,\vartheta\varphi} - \cot\vartheta T_{,\varphi})/\sin\vartheta & (T_{,\varphi\varphi} + \sin\vartheta\cos\vartheta T_{,\vartheta})/\sin^2\vartheta \end{pmatrix},
\end{aligned}
\tag{12}
$$

where for $a, b = \vartheta, \varphi$ we have

$$T_{,ab} := \sum_{l,m} a_{lm} Y_{lm,ab} \quad \text{and} \quad Y_{lm,a} := \frac{\partial}{\partial a} Y_{lm}, \qquad Y_{lm,ab} := \frac{\partial^2}{\partial a\, \partial b} Y_{lm}.$$

Explicit expressions for partial derivatives of the spherical harmonics can be found in Varshalovich, Moskalev and Khersonskii (1988); we have, for instance,

$$\frac{\partial}{\partial \varphi} Y_{lm}(\vartheta, \varphi) = im Y_{lm}(\vartheta, \varphi),$$

$$\frac{\partial}{\partial \vartheta} Y_{lm}(\vartheta, \varphi) = \frac{1}{2}\sqrt{l(l+1) - m(m+1)} Y_{l,m+1}(\vartheta, \varphi) e^{-i\varphi}$$

$$-\frac{1}{2}\sqrt{l(l+1)-m(m-1)}Y_{l,m-1}(\vartheta,\varphi)e^{i\varphi}.$$

In particular, the eigenvalues of the Hessian matrix $H$ are intrinsic quantities and do not depend on the choice of coordinates; hence, points can be classified as hills, lakes or saddles by checking whether both eigenvalues are positive, both negative or with opposite sign, respectively. The next step in the procedure is to focus on thresholded random fields, that is, to consider only those values where $T > v$, for some level $v$ which is taken between $\pm 3\sigma$ from zero ($\sigma$ denoting as usually the standard deviation of $T$). The relative proportion of any two of the curvature typologies can then be evaluated as a function of a threshold value $\nu$; that is, if we consider hills $h(v)$ and lakes $l(v)$ we get

$$h(v) := \frac{\#\{x_i : \lambda_1(H(x_i)), \lambda_2(H(x_i)) \geq 0, T(x_i) \geq \nu\}}{\#\{x_i : T(x_i) \geq \nu\}},$$

$$l(v) := \frac{\#\{x_i : \lambda_1(H(x_i)), \lambda_2(H(x_i)) < 0, T(x_i) \geq \nu\}}{\#\{x_i : T(x_i) \geq \nu\}},$$

where $\lambda_1(H(x_i)), \lambda_2(H(x_i))$ denote the two eigenvalues of $H(\cdot)$ at the location $x_i$, $\{x_i\}$ denoting any discretization of the sphere as provided, for instance, by *HealPix*. The same procedures can then be evaluated on a grid of different threshold values $\nu_j$, $j = 1, \ldots, q$, and this leads to normalized statistics

$$l'(\nu_j) := \frac{l(\nu_j)}{El(\nu_j)}, \qquad h'(\nu_j) := \frac{h(\nu_j)}{Eh(\nu_j)}.$$

It must be stressed that Dore', Colombi and Bouchet (2003) provided some analytic results for $El(\nu_j), Eh(\nu_j)$ in the standard Euclidean case; as these procedures depend only on local features, these analytic results provide excellent approximations even in the spherical case, as shown in Cabella et al. (2005). On the other hand, currently there is no rigorous result on the asymptotic distribution of such statistics, which must hence be calibrated by simulations.

In Figures 4 and 5 we report the $1\sigma$ confidence regions for the hills and lakes densities at various thresholds, evaluated as before by simulations on 1000 Gaussian random fields which mimics the expected behavior of CMB radiations [see again Cabella et al. (2005) for details]. As in the previous subsection, the dotted and dashed lines represent Monte Carlo expected values for the values of our statistics for $f_{NL} = \pm 100, \ldots, 500$, respectively. We also report our estimates based on the WMAP 2003 (crosses) and 2007 (squares) data releases. The general conclusions seem rather close to what we found for the bispectrum: the evidence for non-Gaussianity is apparently weak. On the other hand, the non-Gaussian simulations seem to suggest that the power here may be slightly weaker than for the bispectrum, and in any case insufficient to detect values of $f_{NL}$ smaller than 100, as predicted by the theorists. Again, the new data releases from *Planck* are mandatory to reach firmer conclusions in this area.

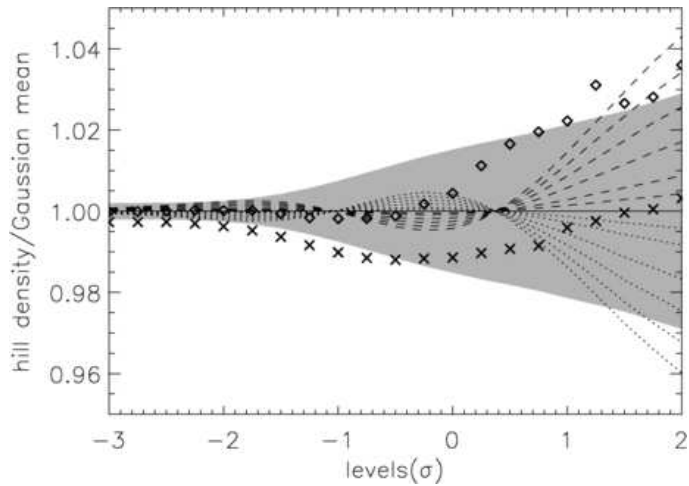

Fig. 4. *"Hills" density on WMAP data.*

As a final remark, we stress that local curvature methods entail a possibility which is ruled out by the bispectrum: as the methods are local, they can be used to test for isotropy, for instance, comparing the behavior of the local curvature on different hemispheres of the CMB sky. This is in-deed the approach pursued by Hansen et al. (2004), where the results on *WMAP* data are compared with Monte Carlo simulations, presenting some boundary evidence that the assumption of isotropy may fail. The literature on the possible existence of these asymmetries has grown enormously in the last 4 years, but no consensus has been reached. As mentioned earlier, the existence of asymmetries would have very profound consequences on cosmological theory, and again, much progress in this area is awaited in the next decade.

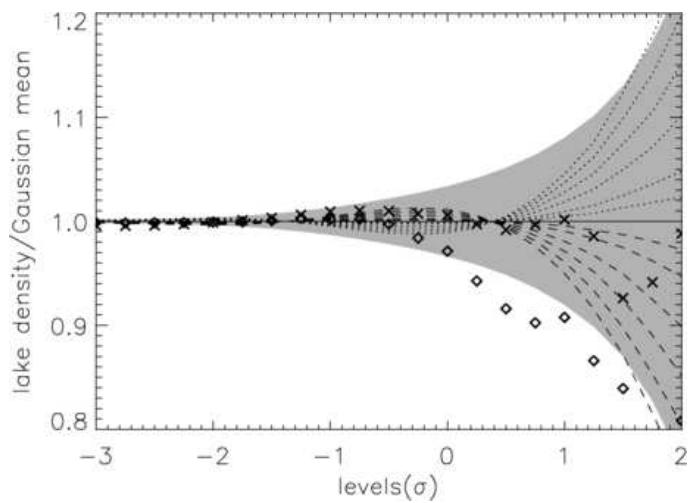

Fig. 5. *"Lakes" density on WMAP data.*

4.3. *Spherical Mexican Hat Wavelets.* Many different proposals have been put forward for the implementation of spherical wavelets systems on the sphere: for the approaches which are more directly connected to the cosmological community, see, for instance, Antoine and Vandergheynst (2007), McEwen et al. (2007), Wiaux, McEwen and Vielva (2007) and the references therein. The construction by Antoine and Vandergheynst (1999), later developed by Wiaux, Jacques and Vandergheynst (2005) has become especially popular; application to CMB data with the aim of testing for possible non-Gaussianity is due to Cruz et al. (2007) and Cruz et al. (2006). We shall focus here on a version of the same approach, that is, the so-called Spherical Mexican Hat Wavelets (hereafter SMHW). The idea of the construction can be explained as follows: in general, a wavelet system on $\mathbb{R}$ can be characterized by means of dilations and translations of a fundamental function (the mother wavelet). On the sphere, the idea is to replace the translations by rotations, that is, elements of the group $SO(3)$. To implement dilations, we note that around the North Pole the latter can be implemented by considering usual dilations in the tangent plane, which are lifted on the sphere by inverse stereographic projections from the South Pole. More precisely, after the identification of the tangent plane at the North Pole with the complex line $\mathbb{C}$, the projection of a point $\omega = (\vartheta, \varphi)$ is provided by $\Phi(\omega) =: \zeta = 2e^{i\varphi} \tan \frac{\vartheta}{2}$. So a stereographic dilation $D_a : S^2 \to S^2$ reads $D_a(\vartheta, \varphi) = (\vartheta_a, \varphi)$, where $\vartheta_a : \tan \frac{\vartheta_a}{2} = a \tan \frac{\vartheta}{2}$, that is, $\vartheta_a := \arctan\{2a \tan \frac{\vartheta}{2}\}$.

More explicitly, the procedure to implement SMHW can be described as follows. In $\mathbb{R}^2$, the continuous Mexican Hat Wavelet can be written as

$$\Psi(x, R) := \frac{1}{\sqrt{2\pi}R} \left[ 2 - \left( \frac{|x|}{R} \right)^2 \right] \exp(-|x|^2/2R),$$

which satisfies the standard compensation and admissibility conditions

$$\int_{R^2} \Psi(x, R) \, dx \equiv 0, \qquad \int_{R^2} \frac{|\widehat{\Psi}(x, R)|^2}{x} \, dx =: C_\Psi < \infty,$$

the hat denoting Fourier transform. For a given scale $R$ and location $x \in S^2$, the definition of the (continuous) Spherical Mexican Hat Wavelet transform can then be entertained in three steps:

- A change of coordinates is performed, to rotate $x$ into the North Pole.
- A stereographic projection on the tangent plane is implemented, so that $y := 2\tan(\frac{\vartheta}{2})$, $\vartheta$ denoting as usual the radial distance from the North Pole. We then implement the MHW on $y$.
- A rotation is performed to transform back the wavelet coefficients to the original location.

It should be noted how the same formalism can be fully justified without the need for stereographic projections, and resorting instead to group representation theory; we refer to Antoine and Vandergheynst (1999) and Wiaux, Jacques and Vandergheynst (2005) for more details. Following this route, the SMHW basis that we implement is given at the North Pole by

$$\Psi(x,R) := \frac{1}{\sqrt{2\pi}N(R)}\left[1+\left(\frac{x}{2}\right)^2\right]^2\left[2-\left(\frac{x}{R}\right)^2\right]\exp(-x^2/2R),$$

with corresponding random coefficients

$$w(R) := \int_{S^2} T(x)\Psi(x,R)\,dx,$$

where $N(R)$ is a normalizing constant and $x$ denotes the polar angle obtained with the stereographic projection; see also Vielva et al. (2003), Cruz et al. (2007) and Cruz et al. (2006). SMHW do not represent a tight frame, so no exact reconstruction formula is available. Their stochastic properties are currently under investigation to establish whether their random coefficients are asymptotically uncorrelated, as it was the case for needlets. On the other hand, SMHW do enjoy very good localization properties [see Marinucci et al.

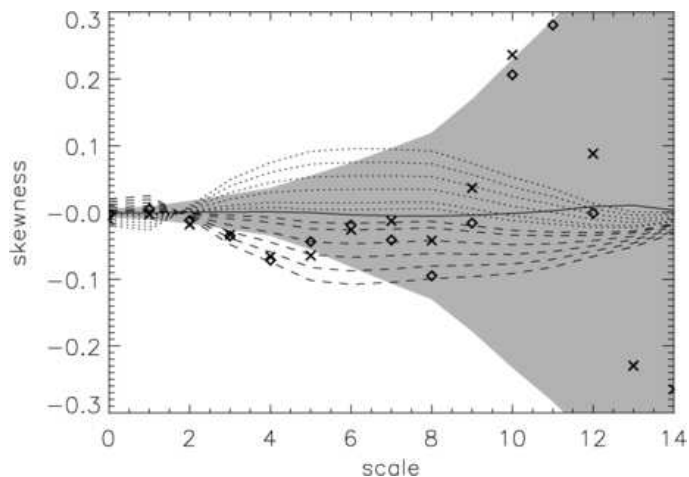

Fig. 6. *The behavior of SMHW on WMAP data.*

(2008) for a comparison with needlets], they are simple to implement and they have been very widely used for Gaussianity and isotropy testing. For completeness, we thus implement the resulting coefficients as a test of non-Gaussianity, to be compared with our previous findings. In particular, we considered skewness and kurtosis for the SMHW random coefficients, which we calibrated by means of Gaussian simulations. For the skewness, the result are reported in Figure 6; for kurtosis, they are not reported, as the resulting power properties where worse. The shaded area, the dotted and dashed lines, the crosses (2003) and the squares (2007) have the same meaning as before: it should be noted that in the $x$-axis is reported the scale factor $R$, so that when moving from left to right we are approaching larger scales (i.e., smaller frequencies, the opposite than for the bispectrum, which is a frequency-domain statistic).

As before, the evidence for non-Gaussianity is weak; worse than that, here simulations suggest that much larger values of $f_{NL}$ would be needed to ensure detection. It seems thus that this class of methods cannot outperform procedures such as the bispectrum when looking for non-Gaussianity. It must be recalled, however, that wavelets do enjoy a very important advantage on pure harmonic methods: indeed, their localization properties in the real domain allow the detection of unexpected features which may signal anisotropic behavior. A striking example of this possibility is provided by Cruz et al. (2007) and Cruz et al. (2006), where a form of SMHW has been used to detect a *Cold Spot* in CMB radiation. The existence and possible explanations for such features are again very widely debated—there seems to be a tight relationship with the evidence on asymmetries which was mentioned earlier [Park (2004) and Hansen et al. (2004)]. This is one more area where new statistical challenges will take place on *Planck* data, and the most suitable forms of spherical wavelets will certainly provide valuable contributions.

## APPENDIX

**A.1. Isotropy and Wigner's coefficients.** A proper derivation of the spherical bispectrum expression would require a considerable effort and some nontrivial background on group representation theory. We report here just the basic facts, and refer to the literature for a more detailed discussion [see, e.g., Varshalovich, Moskalev and Khersonskii (1988), Vilenkin and Klymik (1991), Hu (2001) and Marinucci (2006, 2008)].

The spherical harmonics are defined by

$$Y_{lm}(\theta, \varphi) = \sqrt{\frac{2l+1}{4\pi} \frac{(l-m)!}{(l+m)!}} P_{lm}(\cos\theta) \exp(im\varphi) \qquad \text{for } m > 0,$$

$$Y_{lm}(\theta, \varphi) = (-1)^m \overline{Y}_{l|m|}(\theta, \varphi) \qquad \text{for } m < 0,$$

where $P_{lm}(\cos\theta)$ denotes the associated Legendre polynomial of degree $l, m$, that is,

$$P_{lm}(x) = (-1)^m (1-x^2)^{m/2} \frac{d^m}{dx^m} P_l(x), \qquad P_l(x) = \frac{1}{2^l l!} \frac{d^l}{dx^l}(x^2-1)^l,$$

$$m = 0, 1, 2, \ldots, l, \qquad l = 1, 2, 3, \ldots.$$

Here, $0 \leq \vartheta \leq \pi$ and $0 \leq \vartheta < 2\pi$ denote the usual spherical coordinates on the sphere; more explicit expressions for $Y_{lm}(\cdot)$ are given below [see (14)]. The spherical harmonics provide an orthonormal system for the space of square integrable functions $L^2(S^2)$ with the uniform measure. Now let $g \in SO(3)$ be an arbitrary rotation of $\mathbb{R}^3$ (i.e., a change of coordinate). It is a well-known fact in spherical geometry that the action of the rotation group can be parametrized by the three Euler angles $g = (\alpha, \beta, \gamma)$, $0 \leq \alpha < 2\pi$, $0 \leq \beta \leq \pi$, $0 \leq \gamma < 2\pi$. The action of $SO(3)$ on the spherical harmonics can instead be expressed as

$$(13) \qquad Y_{lm}(gx) \equiv \sum_{m'=-l}^{l} D^l_{m'm}(g) Y_{lm'}(x),$$

where $\{D^l_{m'm}(\cdot)\}$ are the so-called Wigner's $D$ matrices, which provide an irreducible representation of the group of rotations $SO(3)$. In coordinates, an explicit expression for the elements of the $D$-matrices is provided by

$$D^l(\alpha, \beta, \gamma) = \{D^l_{m'm}(\alpha, \beta, \gamma)\}_{m',m=-l,\ldots,l} = \{e^{-im'\alpha} d^l_{m'm}(\beta) e^{im\gamma}\}_{m',m=-l,\ldots,l},$$

where

$$d^l_{mn}(\vartheta) = (-1)^{l-n} [(l+m)!(l-m)!(l+n)!(l-n)!]^{1/2}$$
$$\times \sum_k (-1)^k \frac{(\cos(\vartheta/2))^{m+n+2k} (\sin(\vartheta/2))^{2l-m-n-2k}}{k!(l-m-k)!(l-n-k)!(m+n+k)!},$$

and the sum runs over all $k$ such that the factorials are nonnegative; see Varshalovich, Moskalev and Khersonskii (1988) and Vilenkin and Klymik (1991) for a huge collection of alternative expressions. The proof of (13) is based upon group representation theory techniques and we do not provide it here; we simply recall that the elements of $D^l(\alpha, \beta, \gamma)$ are related to the spherical harmonics by the relationship

$$(14) \quad D^l_{0m}(\alpha, \beta, \gamma) = (-1)^m \sqrt{\frac{4\pi}{2l+1}} Y_{l-m}(\beta, \alpha) = \sqrt{\frac{4\pi}{2l+1}} \overline{Y}_{lm}(\beta, \alpha).$$

By exploiting (13), it is readily seen that isotropy (i.e., rotational invariance in law) entails

$$\sum_{l=0}^{\infty} \sum_{m=-l}^{l} a_{lm} Y_{lm}(x) \stackrel{d}{=} \sum_{l=0}^{\infty} \sum_{m=-l}^{l} a_{lm} Y_{lm}(gx)$$

$$= \sum_{l=0}^{\infty} \sum_{m=-l}^{l} a_{lm} \sum_{m'=-l}^{l} D_{m'm}^{l}(g) Y_{lm'}(x)$$

$$= \sum_{l=0}^{\infty} \sum_{m'=-l}^{l} \left( \sum_{m=-l}^{l} a_{lm} D_{m'm}^{l}(g) \right) Y_{lm'}(x),$$

that is,

(15)  $$(a_{l.}) \stackrel{d}{=} (D^{l}(g) a_{l.}), \qquad l = 1, 2, \ldots,$$

where the identity in law holds jointly over $l$ and $(a_{l.})$ denotes the $2l+1$ vector $(a_{l-m}, \ldots, a_{lm})$. We now recall the expression for the so-called Gaunt integral

(16)
$$\int_{S^2} Y_{l_1 m_1}(x) Y_{l_2 m_2}(x) Y_{l_3 m_3}(x) \, dx$$
$$= \sqrt{\frac{(2l_1+1)(2l_2+1)(2l_3+1)}{4\pi}} \begin{pmatrix} l_1 & l_2 & l_3 \\ 0 & 0 & 0 \end{pmatrix} \begin{pmatrix} l_1 & l_2 & l_3 \\ m_1 & m_2 & m_3 \end{pmatrix}$$

which again requires a group-theoretical proof; indeed, the so-called Wigner's 3j coefficients can be viewed as elements of the unitary matrices which alternate representation of the group of rotations; see Varshalovich, Moskalev and Khersonskii (1988), Vilenkin and Klymik (1991), Marinucci (2006) and Marinucci and Peccati (2007) for a more detailed discussion and explicit expressions.

From (13) we have that under an arbitrary rotation the spherical harmonics transform as

(17)  $$\widetilde{a}_{lm} = \sum_{m'} D_{m'm}^{l}(g) a_{lm'},$$

and

$$E \widetilde{a}_{l_1 m_1} \widetilde{a}_{l_2 m_2} \widetilde{a}_{l_3 m_3} = \sum_{m_1' m_2' m_3'} D_{m_1 m_1'}^{l_1}(g) D_{m_2 m_2'}^{l_2}(g) D_{m_3 m_3'}^{l_3}(g) E a_{l_1 m_1'} a_{l_2 m_2'} a_{l_3 m_3'}$$

$$= \sum_{m_1' m_2' m_3'} D_{m_1 m_1'}^{l_1}(g) D_{m_2 m_2'}^{l_2}(g) D_{m_3 m_3'}^{l_3}(g)$$
$$\times \begin{pmatrix} l_1 & l_2 & l_3 \\ 0 & 0 & 0 \end{pmatrix} \begin{pmatrix} l_1 & l_2 & l_3 \\ m_1' & m_2' & m_3' \end{pmatrix} b_{l_1 l_2 l_3}$$
$$= \begin{pmatrix} l_1 & l_2 & l_3 \\ 0 & 0 & 0 \end{pmatrix} \begin{pmatrix} l_1 & l_2 & l_3 \\ m_1 & m_2 & m_3 \end{pmatrix} b_{l_1 l_2 l_3},$$

where we have used the identity

$$\sum_{m_1' m_2' m_3'} \begin{pmatrix} l_1 & l_2 & l_3 \\ m_1' & m_2' & m_3' \end{pmatrix} D_{m_1 m_1'}^{l_1}(g) D_{m_2 m_2'}^{l_2}(g) D_{m_3 m_3'}^{l_3}(g)$$

(18)
$$\equiv \begin{pmatrix} l_1 & l_2 & l_3 \\ m_1 & m_2 & m_3 \end{pmatrix}.$$

For a proof of (18), it suffices to use (16) and note that

$$\sum_{m_1' m_2' m_3'} \begin{pmatrix} l_1 & l_2 & l_3 \\ m_1' & m_2' & m_3' \end{pmatrix} D^{l_1}_{m_1 m_1'}(g) D^{l_2}_{m_2 m_2'}(g) D^{l_3}_{m_3 m_3'}(g)$$

$$= \left\{ \begin{pmatrix} l_1 & l_2 & l_3 \\ 0 & 0 & 0 \end{pmatrix} \sqrt{\frac{(2l_1+1)(2l_2+1)(2l_3+1)}{4\pi}} \right\}^{-1}$$

$$\times \int_{S^2} \left[ \sum_{m_1' m_2' m_3'} D^{l_1}_{m_1 m_1'}(g) D^{l_2}_{m_2 m_2'}(g) D^{l_3}_{m_3 m_3'}(g) Y_{l_1 m_1'} Y_{l_2 m_2'} Y_{l_3 m_3'} \right] dx$$

$$= \left\{ \begin{pmatrix} l_1 & l_2 & l_3 \\ 0 & 0 & 0 \end{pmatrix} \sqrt{\frac{(2l_1+1)(2l_2+1)(2l_3+1)}{4\pi}} \right\}^{-1} \int_{S^2} [Y_{l_1 m_1} Y_{l_2 m_2} Y_{l_3 m_3}] dx$$

$$= \begin{pmatrix} l_1 & l_2 & l_3 \\ m_1 & m_2 & m_3 \end{pmatrix}.$$

For the first step to be justified, we need to ensure the Wigner's $3j$ coefficients within the curly brackets is indeed different from zero; this condition is fulfilled provided $l_1 + l_2 + l_3 = even$ and the so-called triangle conditions are met, namely, $l_i + l_j \leq l_k$ for all choices of $i, j, k = 1, 2, 3$.

We now turn to the issue of sample estimators. We can indeed show immediately that (4) is invariant to rotations; we have, in fact,

$$\frac{1}{2l+1} \sum_{m=-l}^{l} |\widetilde{a}_{lm}|^2 = \frac{1}{2l+1} \sum_{m=-l}^{l} \left| \sum_{m'} D^l_{mm'}(g) a_{lm'} \right|^2$$

$$= \frac{1}{2l+1} \sum_{m_1, m_2} a_{lm_1} \overline{a_{lm_2}} \sum_{m=-l}^{l} D^l_{mm_1}(g) \overline{D^l_{mm_2}(g)}$$

$$= \frac{1}{2l+1} \sum_{m_1, m_2} a_{lm_1} \overline{a_{lm_2}} \delta^{m_2}_{m_1} = \frac{1}{2l+1} \sum_{m=-l}^{l} |a_{lm}|^2,$$

in view of the orthonormality property Varshalovich, Moskalev and Khersonskii (1988)

$$\sum_{m=-l}^{l} D^l_{mm_1}(g) \overline{D^l_{mm_2}(g)} \equiv \delta^{m_2}_{m_1} \delta^{l_2}_{l_1}.$$

A similar argument exploiting (18) and (17) shows indeed that the sample bispectrum is itself invariant to rotations. We refer again to Hu (2001), Marinucci (2006) and Marinucci (2008) for a much wider discussion and more properties.

### A.2. Some background on differential geometry.

A.2.1. *Scalars, vectors and tensors.* Our purpose is to establish intrinsic measures of the local curvature of a random field. Here, by intrinsic we mean "independent from the choice of coordinates," much the same way as our bispectrum statistics in the previous section. To clarify this issue, we recall some basic definition from differential geometry on the sphere; there exist many beautiful books on this issue [Adler and Taylor (2007), Bishop and Goldberg (1980)], and we refer to these for a proper account: we just report some basic facts to make the local curvature approach intuitive. Let $M$ be a smooth manifold and write $\mathbb{T}_x$ for the tangent plane at $x \in M$; we label $\mathbb{T}_x^*$ the cotangent plane, that is, the dual space of $\mathbb{T}_x$ [refer again to Adler and Taylor (2007) and Bishop and Goldberg (1980) for details and definitions]. Let $\phi: M \to \mathbb{R}$ be some smooth function on the manifold; we say $\phi$ is a *scalar function* if it transforms under a change of coordinate $g$ as

$$\overline{\phi}(gx) = \phi(x) \qquad \forall x \in M, \text{ that is, } \overline{\phi} := \phi \circ h^{-1}.$$

A rank one *covariant tensor* is a linear operator on the vector space $\mathbb{T}_x$ with components $\{f_i(\cdot)\}$ which transform according to the rule

$$\overline{f}_j(y) = \sum_i f_i(x) \frac{\partial x^i}{\partial y^j},$$

where we wrote $y := \{y^1, \ldots, y^n\} = \{g_1(x), \ldots, g_n(x)\}$. Likewise, a rank one *contravariant tensor* of dimension $n$ is a linear operator $\{f_i(\cdot)\}_{i=1,\ldots,n}$ whose components transform according to the rule

$$\overline{f}^j(y) = \sum_{i=1}^n f^i(x) \frac{\partial y^j}{\partial x^i},$$

where we wrote as before $y = gx$.

It is then possible to extend this definition to higher orders—for our purposes, rank 2 suffices. A rank two *covariant tensor* of dimension $n$ is a bilinear operator $\{T_{uv}\}_{u,v=1,\ldots,n}$ whose components transform according to the rule

$$\overline{T}_{uv}(y) = \sum_{p,q=1}^n T_{pq}(x) \frac{\partial x^p}{\partial y^u} \frac{\partial x^q}{\partial y^v}. \tag{19}$$

Likewise, we can define contravariant and mixed rank two vectors, denoted by $T^{uv}$ and $T_v^u$ respectively. It is immediately seen from (19) that the usual Hessian matrix of a scalar function is a rank two covariant tensor.

The previous concepts assume a nontrivial meaning in the presence of manifolds with a nonzero intrinsic curvature, such as the sphere. In such spaces, we can introduce the metric tensor $\{g_{ab}(\cdot)\}$ by imposing that the length of a vector $X := \{x^1, \ldots, x^n\}$ be given by

$$\|X\|_g^2 := \sum_{a,b=1}^n g_{ab}(x) x^a x^b.$$

Thus, $\{g_{ab}(x)\}_{a,b=1,\ldots,n}$ is a rank two tensor. Some examples:

- Euclidean case: we have

$$\|X\|_E^2 := \sum_{a,b=1}^n x^a x^b \delta_a^b \qquad \text{that is, } g_{ab} \equiv \delta_a^b.$$

- The two sphere $S^2$: in spherical coordinates, for $\Omega := \{\vartheta, \varphi\}$ the usual metric is given by

(20) $$\{g_{ab}(\Omega)\}_{a,b=\vartheta,\varphi} = \begin{pmatrix} 1 & 0 \\ 0 & \sin^2 \vartheta \end{pmatrix}.$$

The contravariant or reciprocal metric tensor $\{g^{ab}(x)\}_{a,b=1,\ldots,n}$ is defined by the requirement that

$$\sum_b g^{ab}(x) g_{bc}(x) \equiv \delta_c^a,$$

that is, it represents the elements of the inverse matrix $g^{-1}$. By means of this tensor operator, we can define more generally contravariant vectors, denoting, for instance,

$$T_a := \sum_b g_{ab} T^b \quad \text{and} \quad \text{consequently,} \quad T^a = \sum_b g^{ab} T_b,$$

so that we ensure invariance, that is,

$$\|\{T_a\}_{a=1,\ldots,n}\|_g = \sum_a T^a T_a = \sum_{a,b} g_{ab} T^a T^b = \sum_{a,b} g^{ab} T_a T_b =: \|\{T^a\}_{a=1,\ldots,n}\|_g.$$

Covariant tensors can be likewise introduced, that is,

$$T_b^a = \sum_c g^{ac} T_{ac}, \qquad T^{ab} = \sum_c g^{ac} T_c^b = \sum_{c,d} g^{ac} g^{bd} T_{cd}.$$

Let us now investigate the behavior of second order derivatives under coordinate transformations. We have

$$\frac{\partial^2 \overline{\phi}}{\partial y^u \, \partial y^v} = \frac{\partial}{\partial y^u} \left( \frac{\partial \overline{\phi}}{\partial y^v} \right)$$

$$= \frac{\partial}{\partial y^u} \left( \sum_{p=1}^n \phi_p(x) \frac{\partial x^p}{\partial y^v} \right)$$

$$= \sum_{p,q=1}^n \phi_{pq}(x) \frac{\partial x^p}{\partial y^v} \frac{\partial x^q}{\partial y^u} + \sum_{p=1}^n \phi_p(x) \frac{\partial^2 x^p}{\partial y^u \, \partial y^v}.$$

It is therefore clear that for nonlinear coordinate transformations (i.e., such that $\{\partial^2 x^p/\partial y^u \partial y^v\} \neq 0$) standard second order derivatives do not act as a rank two tensor, that is, they depend on the coordinate choice and do not represent intrinsic quantities. To overcome this problem, we need to introduce the well-known Christoffel coefficients $\Gamma^l_{ik}$, which, by assumption, satisfy the transformation laws

$$\overline{\Gamma}^l_{ik} = \sum_{m,n,j} \Gamma^n_{jm} \frac{\partial x^j}{\partial y^i} \frac{\partial x^m}{\partial y^k} \frac{\partial y^l}{\partial x^n} + \sum_j \frac{\partial^2 x^j}{\partial y^i \partial y^k} \frac{\partial y^l}{\partial x^j}.$$

It is then easy to check that

$$\sum_l \overline{\Gamma}^l_{ik} \overline{\phi}_l = \sum_{m,n,j} \Gamma^n_{jm} \phi_n \frac{\partial x^j}{\partial y^i} \frac{\partial x^m}{\partial y^k} + \sum_j \phi_j \frac{\partial^2 x^j}{\partial y^i \partial y^k}$$

and, hence,

$$\overline{\phi}_{i,k} - \sum_l \overline{\Gamma}^l_{ik} (L_h \phi)_l = \sum_{m,j} \left( \phi_{j,m} - \sum_n \Gamma^n_{jm} \phi_n \right) \frac{\partial x^j}{\partial y^i} \frac{\partial x^m}{\partial y^k},$$

that is, $\{\phi_{j,m} - \sum_n \Gamma^n_{jm} \phi_n\}$ is actually a rank two tensor. The previous discussion motivates the following:

DEFINITION. The covariant derivative of the rank one tensor $T_i$ is given by

$$T_{i;k} := T_{i,k} - \sum_l \Gamma^l_{ik} T_l,$$

where $T_{i,k} = \partial T_i/\partial x^k$ denotes standard derivative.

Let us now specify the previous definition to the sphere. In terms of the metric tensor, the Christoffel symbols can be shown to be equal to

(21) $$\Gamma^l_{ij} = \sum_k \left\{ \frac{g^{kl}}{2}(g_{ki,j} + g_{kj,i} - g_{ij,k}) \right\}.$$

For instance, in the Euclidean case $g_{ki,j} = g_{kj,i} = g_{ij,k} \equiv 0$, hence, $\Gamma^l_{ij} \equiv 0$; thus, covariant derivatives coincide with standard calculus operators. On the contrary, for the sphere $S^2$, we have $g_{\vartheta\vartheta} = g^{\vartheta\vartheta} = 1$, $g_{\varphi\varphi} = \sin^2\vartheta = (g^{\varphi\varphi})^{-1}$, $g_{\vartheta\varphi} = g^{\vartheta\varphi} = 0$. Hence, we obtain

$$g_{\vartheta\vartheta,\vartheta} = g_{\vartheta\vartheta,\varphi} = g_{\vartheta\varphi,\vartheta} = g_{\vartheta\varphi,\varphi} = g_{\varphi\varphi,\varphi} = 0$$

and $g_{\varphi\varphi,\vartheta} = 2\sin\vartheta\cos\vartheta$; to summarize, the Christoffel symbols on the sphere are

$$\Gamma^{\vartheta}_{\vartheta\vartheta} = \Gamma^{\varphi}_{\vartheta\vartheta} = 0,$$

$$\Gamma^{\varphi}_{\varphi\vartheta} = \frac{g^{\varphi\varphi}}{2}\{g_{\varphi\varphi,\vartheta} + g_{\varphi\vartheta,\varphi} - g_{\varphi\vartheta,\varphi}\} = \frac{1}{2\sin^2\vartheta}2\sin\vartheta\cos\vartheta = \cot\vartheta,$$

$$\Gamma^{\vartheta}_{\varphi\vartheta} = \frac{g^{\vartheta\vartheta}}{2}\{g_{\varphi\vartheta,\vartheta} + g_{\vartheta\vartheta,\varphi} - g_{\varphi\vartheta,\vartheta}\} + \frac{g^{\vartheta\varphi}}{2}\{g_{\varphi\varphi,\vartheta} + g_{\vartheta\varphi,\varphi} - g_{\varphi\vartheta,\varphi}\} = 0,$$

$$\Gamma^{\varphi}_{\varphi\varphi} = \frac{g^{\varphi\varphi}}{2}g_{\varphi\varphi,\varphi} = 0,$$

$$\Gamma^{\vartheta}_{\varphi\varphi} = \frac{g^{\vartheta\vartheta}}{2}\{g_{\varphi\vartheta,\varphi} + g_{\varphi\vartheta,\varphi} - g_{\varphi\varphi,\vartheta}\} = -\sin\vartheta\cos\vartheta.$$

Hence, we obtain the following results for the covariant derivatives of the spherical harmonics:

$$Y_{lm;\vartheta\vartheta} \equiv Y_{lm,\vartheta\vartheta},$$

$$Y_{lm;\vartheta\varphi} \equiv Y_{lm,\vartheta\varphi} - \Gamma^{\varphi}_{\varphi\vartheta}Y_{lm,\varphi} - \Gamma^{\vartheta}_{\varphi\vartheta}Y_{lm,\vartheta} = Y_{lm,\vartheta\varphi} - \cot\vartheta Y_{lm,\varphi},$$

$$Y_{lm;\varphi\varphi} \equiv Y_{lm,\varphi\varphi} - \Gamma^{\varphi}_{\varphi\varphi}Y_{lm,\varphi} - \Gamma^{\vartheta}_{\varphi\varphi}Y_{lm,\vartheta} = Y_{lm,\varphi\varphi} + \sin\vartheta\cos\vartheta Y_{lm,\vartheta}.$$

The previous expressions provide the clue for the computation of the bilinear form

$$(22) \qquad H^* := \begin{pmatrix} T_{;\vartheta\vartheta} & T_{;\vartheta\varphi} \\ T_{;\vartheta\varphi} & T_{;\varphi\varphi} \end{pmatrix} = \begin{pmatrix} \sum_{lm} a_{lm}Y_{lm;\vartheta\vartheta} & \sum_{lm} a_{lm}Y_{lm;\vartheta\varphi} \\ \sum_{lm} a_{lm}Y_{lm;\vartheta\varphi} & \sum_{lm} a_{lm}Y_{lm;\varphi\varphi} \end{pmatrix}.$$

To obtain (12), we need to introduce a final, quite subtle point. (22) defines a bilinear form $H^*:(\mathbb{T}\times\mathbb{T})\to\mathbb{R}$ acting on the tensor product of the tangent plane with itself; in order to be able to evaluate consistently the eigenvalues, we must transform this form into the corresponding linear operator $H:\mathbb{T}\to\mathbb{T}$, where $H := g^{-1}H^*$ [actually we considered the symmetrized expression $H := g^{-1/2}H^*g^{-1/2}$, where $g$ denotes as before the metric tensor on the sphere, see (20)]. This explains the appearance of the $\sin\vartheta$ factors at the denominators in (12)—we refer again to Bishop and Goldberg (1980) for more details and explanations.

**Acknowledgment.** We are grateful to Frode K. Hansen for a long standing collaboration in this area.

DEPARTMENT OF PHYSICS  
UNIVERSITY OF ROME TOR VERGATA  
VIA DELLA RICERCA SCIENTIFICA 1  
00133 ROMA  
ITALY  
E-MAIL: paolo.cabella@roma2.infn.it

DEPARTMENT OF MATHEMATICS  
UNIVERSITY OF ROME TOR VERGATA  
VIA DELLA RICERCA SCIENTIFICA 1  
00133 ROMA  
ITALY  
E-MAIL: marinucc@mat.uniroma2.it